**Democratizing Advanced High-Throughput Imaging via Cross-Instrument Deep Learning-Enabled Modality Transfer**

Dominik Panek[a, b]*, Carina Rząca[a, b], Maksymilian Szczypior[b,c], Joanna Sorysz[d,#], Krzysztof Misztal[e], Zbigniew Baster[b,f]*, Zenon Rajfur[b]*

[a] Doctoral School of Exact and Natural Sciences, Jagiellonian University, ul. Łojasiewicza 11, 30-348 Kraków, Poland.

[b] Department of Molecular and Interfacial Biophysics, Faculty of Physics, Astronomy and Applied Computer Science, Jagiellonian University, ul. Łojasiewicza 11, 30-348 Kraków, Poland.

[c] Undergraduate program in Biophysics, Faculty of Physics, Astronomy and Applied Computer Science, Jagiellonian University, ul. Łojasiewicza 11, 30-348 Kraków, Poland.

[d] Undergraduate program in Biomedical Engineering, Faculty of Electrical Engineering, Automatics, Computer Science and Biomedical Engineering, AGH University of Science and Technology, Al. A. Mickiewicza 30, Building B-1, 30-059 Kraków, Poland.

[e] Division of Computational Mathematics, Faculty of Mathematics and Computer Science, Jagiellonian University, 6 Łojasiewicza St., Kraków, Poland.

[f] Laboratory for Cell and Tissue Engineering, Department of Biomedical Engineering, Eindhoven University of Technology, 5600 MB Eindhoven, The Netherlands.

[#] Current address: German Research Center for Artificial Intelligence (DFKI) and RPTU Kaiserslautern-Landau, Kaiserslautern, Germany.

*Authors to whom correspondence should be addressed: dominik.panek@doctoral.uj.edu.pl, z.j.baster@tue.nl, zenon.rajfur@uj.edu.pl.

**Abstract**

High-throughput imaging is often constrained by a trade-off between acquisition speed and image quality. Fast imaging modalities, such as wide-field fluorescence microscopy, enable large-scale data acquisition but suffer from reduced contrast and resolution, whereas high-resolution techniques, like confocal or super-resolution techniques, provide superior image quality at the cost of reduced throughput and increased instrument time.

Here, we present a deep learning-based approach for modality transfer across independent microscopes, enabling the transformation of low-quality images acquired on fast systems into high-quality representations comparable to those obtained using advanced imaging platforms. To achieve this, we employed a generative adversarial network (GAN)-based model trained on paired datasets acquired on physically separate wide-field and confocal microscopes, demonstrating that image quality can be reliably transferred between independent instruments. Quantitative evaluation shows substantial improvement in structural similarity and signal fidelity, with median SSIM and PSNR values of 0.94 and 31.87, respectively, compared to 0.83 and 21.48 for the original wide-field images. These results indicate that key structural features can be recovered with high accuracy.

Importantly, this approach enables a workflow in which high-throughput imaging can be performed on fast, accessible microscopy systems while preserving the ability to computationally recover high-quality structural information. High-resolution microscopy can then be reserved for algorithm training and targeted validation, reducing acquisition time and improving overall experimental efficiency. This workflow supports a model in which shared imaging facilities provide access to advanced instrumentation without requiring individual research groups to procure dedicated high-end systems.

Together, our results establish deep learning-enabled modality transfer as a practical strategy for bridging independent microscopy systems and supporting scalable, high-content imaging workflows.

## 1. Introduction

High-quality microscopy plays a crucial role in biological sciences [1,2]. The development of several groundbreaking techniques, such as fluorescence confocal microscopy and later super-resolution methods, has unlocked new possibilities in biological research, including 3D optical sectioning and imaging nanometric-sized (5 – 20 nm) structures [3–8]. Such imaging provides quantitative information that can be used to model structures and processes at the subcellular level [9–11]. Nowadays, these advanced techniques are well established in the spectrum of research tools, and it is difficult to imagine modern biological research without them. However, despite continued technological progress, they remain associated with limitations such as high equipment cost and relatively low imaging throughput, which restrict their accessibility and scalability [4,12].

This limitation is particularly relevant in high-throughput imaging (HTI) applications, where experiments involve acquiring large numbers of images across multiple samples or imaging locations [13]. At such scales, cumulative acquisition time becomes a substantial practical constraint. Consequently, throughput and system accessibility favor simpler imaging modalities, limiting the routine use of high-resolution techniques in large-scale experiments [14].

In this context, techniques such as basic wide-field fluorescence (WFF) microscopy have become increasingly relevant for HTI due to their accessibility and high acquisition speed. Furthermore, second-hand or retired advanced systems, available at a fraction of their original cost, can still provide robust fluorescence imaging capabilities [15,16]. These systems enable fast, cost-effective image acquisition, making them well-suited for large-scale experiments, but lack the optical sectioning and resolution provided by more advanced techniques. This reinforces the fundamental trade-off between imaging throughput and image quality in modern microscopy.

Various approaches have been developed to improve image quality or to transform images from one imaging modality into another, through computational processing [17,18]. In recent years, deep learning methods, particularly generative adversarial networks (GANs), have emerged as powerful tools for image enhancement and transformation. GANs consist of two competing networks, a generator and a discriminator, that are trained in an adversarial manner [19,20]. This framework has been successfully applied to a range of image restoration and super-resolution tasks [21–26], including cases where images from the same imaging system are acquired at different resolutions or under different imaging conditions [27]. Moreover, various modifications of GAN architectures have been shown to be effective across a range of image enhancement applications [21,26–29]. A key advantage of GAN-based approaches is their flexibility, which allows straightforward integration of architectural modifications. For example, the incorporation of Fourier Channel Attention blocks has been shown to improve performance in tasks involving STED and confocal image comparison [30]. However, increasing model complexity typically results in higher computational demands, which can significantly extend processing time.

In this study, we apply deep learning-based image enhancement to transfer imaging modality between independent microscopy systems: wide field and scanning confocal. To that, we developed a GAN-based model to recover high-quality structural information in low-quality WFF images using their confocal counterparts as ground truth. The generator is trained to produce high-quality representations from low-quality inputs, while the discriminator evaluates their similarity to experimentally acquired confocal images [19]. For the generator, we employed a U-NET architecture [31]. Historically, U-NET was first applied to biological image segmentation, but has also proven to be a robust architecture for our case due to its ability to efficiently capture multiscale features through down- and up-sampling pathways [32–

[37]. The architecture was further modified with residual connections to improve performance. For the discriminator, we employed a simple convolutional neural network (CNN) architecture.

The results obtained using this cross-system dataset demonstrate that image quality can be transferred between independent microscopy systems. In contrast to prior work focusing primarily on transformations within a single imaging platform, our approach explicitly addresses the case of physically separate instruments. While demonstrated here for wide-field and confocal microscopy, the proposed framework is expected to be applicable to other combinations of imaging modalities. This establishes a framework for combining high-throughput image acquisition with computational transfer of information from an advanced microscopy modality. By reducing need on repeated acquisition with slower or less accessible instruments, the approach enables more efficient use of advanced microscopy systems and facilitate scalable workflows for high-content imaging. Importantly, this workflow supports a model in which shared imaging facilities play a key role in providing access to such advanced instrumentation without requiring individual groups to procure and maintain dedicated high-end systems.

## 2. Materials and Methods

### 2.1. Cell culture and staining

Mouse embryonic fibroblasts (MEF NIH/3T3 - ATCC CRL-1658, ATCC, Manassas, VA, USA) were cultured in DMEM Low Glucose medium (L0066, Biowest, Nuaillé, France) supplemented with 10% FBS (fetal bovine serum; 10270106, Gibco, Thermo Fisher Scientific, Waltham, MA, USA), and 100 I.U./mL Penicillin and 100 µg/mL Streptomycin Solution (L0022, Biowest, Nuaillé, France) at 37°C, 5% CO2, and 100% humidity. Cells were passaged every 2-3 days when confluence reached about 75%. After at least three passages after thawing, 20,000 cells were plated on #1.5 20 mm round coverslips in a six-well plate and cultured with 2 mL of fresh medium for 24 hours before fixation.

For imaging, cells were fixed with 4% paraformaldehyde solution (47608, Sigma-Aldrich, St. Louis, MO, USA) in PBS for 20 min at 37°C, followed by incubation with 0.5% Triton X-100 (T8787, Sigma-Aldrich, St. Louis, MO, USA) in PBS solution for 5 min at RT, and then 0.1 M glycine PBS solution for 10 min in RT with gentle rocking. Then, the samples were rinsed several times with PBS. After rinsing, samples were blocked with a 4% BSA (bovine serum albumin; A7906, Sigma-Aldrich, St. Louis, MO, USA) solution in PBS-T (PBS - 0.05 % Tween 20 (5037, Sigma-Aldrich, St. Louis, MO, USA)) for 1 hour at RT.

Immunofluorescence antibody staining was performed to label microtubules. Cells were incubated overnight with anti-α-Tubulin antibody (1:500 in 4% BSA PBS buffer; T9026, Sigma-Aldrich, St. Louis, MO, USA). Then, they were washed 3 times with PBS-T (5 min, RT), and stained with Alexa Fluor 555-labeled goat anti-mouse IgG (H+L) antibody (1:100 in 4% BSA PBS buffer; A21422, Thermo Fisher Scientific, Waltham, MA, USA) for 1 hour at RT. Then, the samples were washed several times with PBS.

Thereafter, actin and nucleus were stained using PBS solution of Alexa Fluor™ 488 Phalloidin (1:200; A12379, Thermo Fisher Scientific, Waltham, MA, USA), and Hoechst 33342 (2 µg/ml; H3570, Thermo Fisher Scientific, Waltham, MA, USA), respectively, for 30 min at RT. Then, the samples were washed several times with PBS. Before imaging, samples were mounted on the Correscopy sample holder (Correscopy, Krakow, Poland).

### 2.2. Wide-field fluorescence and confocal images gathering

Two microscopes were used: a laser scanning confocal (Zeiss LSM 710 set on Zeiss Axio Observer Z1 body, oil-immersion Plan-Apochromat 40x NA 1.4 objective lens, and ZEN black version 8.10.484 software) and a wide-field fluorescence microscopes (Zeiss Axio Observer Z1 equipped with a Hamamatsu ORCA-Flash4.0 V2 camera, oil-immersion EC Plan-Neofluar 40x NA 1.30 Ph3 objective lens, and ZEN blue version 2.3 software). Correlative imaging was aided using a Correscopy system (Correscopy, Kraków, Poland), which facilitated locating the field of view between microscopes and initial alignment. Calibrations of the Correscopy system were performed for each microscope before every experiment.

Calibration was performed using dedicated Correscopy software (Correscopy, Kraków, Poland). First, a dedicated reference point was identified, and its orientation was aligned with both the sample holder and the microscope. After calibration was recorded, cells were imaged using 40× oil-immersion objectives, with voxel size preserved between the two microscopes. Images were first acquired on the wide-field fluorescence microscope, after which the sample was transferred to the confocal microscope and calibrated for the second imaging step. Wide-field images were acquired at 2048 × 2048 pixels, whereas confocal images were acquired at 1114 × 1114 pixels. To further improve image alignment quality, 200 nm red fluorescent polystyrene beads (F8763, Thermo Fisher Scientific, Waltham, MA, USA) were added during the paraformaldehyde fixation step. Additionally, calibration slides containing 100 nm polystyrene TetraSpeck beads (T7279, Thermo Fisher Scientific, Waltham, MA, USA) were prepared to correct for chromatic aberrations.

### 2.3. Data preprocessing

Prior to model training, image pairs were preprocessed to correct for spatial misalignment between modalities using ImageJ [38–40]. Misalignments included rotation and translational offsets between wide-field and confocal images. The alignment procedure consisted of the following steps:

a. **Rotation correction**: The relative rotation between image pairs was estimated using the angle measurement tool in Fiji, followed by rotation of one image to achieve alignment.
b. **Region extraction**: Corresponding regions of interest were defined by cropping the wide-field images to match the field of view and dimensions of the confocal images.
c. **Translational offset estimation**: Lateral shifts between image pairs were evaluated by selecting corresponding slices from both datasets, merging them into a single stack, and visually assessing overlap using pixel intensity averaging.
d. **Translation correction**: Identified translational offsets were corrected using the Fiji "Translate" function to minimize spatial discrepancies between modalities.
e. **Z-axis alignment**: Corresponding axial ranges were identified by matching the first and last slices of the image stacks across modalities to ensure consistent depth representation.

### 2.4. Deep Learning Network Architecture

In our research, we employed a GAN architecture. It consists of a discriminator and a generator that work adversarially to produce images that resemble those from a high-quality modality. The original data comprised 149 aligned images of mouse embryonic fibroblasts obtained using laser-scanning confocal and wide-field fluorescence microscopes. Data was split into training, testing, and validation subsets (80%, 19%, and 1%, respectively) on the level of original images. Additionally, the training subset was artificially augmented. The augmentation was done using random flips, random rotations, and/or random translations algorithms (all actions applied simultaneously to LQ and HQ ground-truth images). These procedures allowed us to increase the total number of images to approximately 6,000. The training procedure is described below. The architecture of the network is shown in **Figure 1.** The Discriminator comprises a CNN with 10 convolutional blocks. It initiates with a convolutional layer, and its output is fed through subsequent convolutional blocks, where a series of essential operations unfold as follows:

$$x_k = LReLU[Conv2D(LReLU[Conv2D(LReLU[Conv2D(LReLU[Conv2D(x_{k-1})])])])], \quad k = 1, 2, \dots, 10, \qquad \textbf{(1)}$$

where $x_k$ is the output of the *x-th* convolutional block, $x_0$ is the input from the first convolutional layer, *LReLU* is a Leaky Rectified Linear Unit activation function with a slope $a= 0.1$, and *Conv2D* is a two-dimensional convolution operation. *LReLU* is defined as:

$$LReLU(x, a) = \max(0, x) - a \cdot \max(0, -x). \quad \textbf{(2)}$$

After the 10[th] convolutional block, the average pooling layer is inserted to reduce dimensionality, implementing a 4x4 pool size. Subsequently, two fully connected layers (FC) are added, and the network outputs the probability estimated by the sigmoid activation function.

The generative model is based on a residual U-NET architecture, facilitating the acquisition of spatial information by the network. In this case, U-NET consists of 4 down-sampling and four up-sampling blocks, connected by skip layers. Each down-sampling block consists of 3 convolutional blocks, where the following operations are performed:

$$d_k = d_{k-1} + LReLU[Conv2D(LReLU[Conv(LReLU[Conv2D(d_{k-1})])])], \quad \textbf{(3)}$$
$$k = 1, 2, 3, 4.$$

Following the addition operation, an average pooling layer is inserted with a 2x2 pool size. In this case, $d_0$ is the input image. Similarly, to the down-sampling part, the up-sampling blocks consist of three convolutional blocks with the following operations:

$$u_k = LReLU[Conv2D(LReLU[Conv2D(LReLU[Conv2D(Concat\{d_{5-k}, u_{k-1}\})])])], \quad \textbf{(4)}$$
$$k = 1, 2, 3, 4$$

After the last block, a convolution layer, together with the *LReLU* activation function, is added. Here, $u_0$ denotes the output of the layer, which is placed at the bottom of the U-NET. The *Concat* operation denotes the concatenation of the down-sampling and up-sampling layers (so-called skipped connections), which helps improve network performance.

The loss function employed in the discriminative model is a simple binary cross-entropy (BCE). Its primary objective is to determine whether the received images correspond to the HQ ground truth or are generated by the network, effectively acting as a binary classifier. BCE is defined as follows:

$$\mathcal{L}_{D|G}(X, Y) = BCE\big(D\big(G(X)\big) - D(Y), y\big) + BCE\big(D(Y) - D\big(G(X)\big), 1 - y\big), \quad \textbf{(5)}$$

where *X* and *Y* denote the LQ and HQ input images, respectively. *G(·)* is the generative model output, *D(·)* represents the discriminator model prediction, and *y* is the label set during the training (0 when the discriminator is trained). In the case of the generative model, the combined losses of MSE (mean squared error), SSIM (structural similarity index) [41], and BCE are used:

$$\mathcal{L}_{D|G}(X, Y) = \alpha \cdot MSE(G(X), Y) + \beta \cdot (1 - SSIM(G(X), Y)) + \gamma \cdot BCE(\mathrm{D}(\mathrm{G}(\mathrm{X})) - \mathrm{D}(\mathrm{Y}), \mathrm{y}) \quad \textbf{(6)}$$

Here $\alpha$, $\beta$, and $\gamma$ denote the weights given to the respective losses. MSE loss was used to ensure the general fidelity of the generated image. However, the MSE loss cannot be too high relative to SSIM and BCE, because the differences between the HQ ground truth and the LQ images, as measured by MSE, are initially small. SSIM loss, on the other hand, ensures improved visual quality of the generated image, whereas BCE encourages the generator to produce images similar to the HQ ground-truth images. Altogether, the combined weights of MSE and SSIM accounted for approximately 10% of the total loss (for details, see ablation studies **Supplementary Note 1**).

To compare our network's ability to produce HQ images, we also performed deconvolution on LQ images. The deconvolution workflow was done using Fiji ImageJ software. For that purpose, we used

BatchDeconvolution, PSF generator, and the DeconvolutionLab2 plugins [42–45]. PSF was generated according to the Born-Wolf model [44], with the parameters set to be identical to those used during the experiments, i.e., NA = 1.4, immersion refractive index = 1.515, pixel size = 159 nm and the wavelengths identical to those of the maxima of emission spectra of the fluorophores (461 nm, 520 nm, and 565 nm, for Hoechst 33342, Alexa Fluor 488, and Alexa Fluor 555, respectively). The deconvolution was performed separately for all the channels and was done using the Richardson-Lucy algorithm with 10 iterations [46,47].

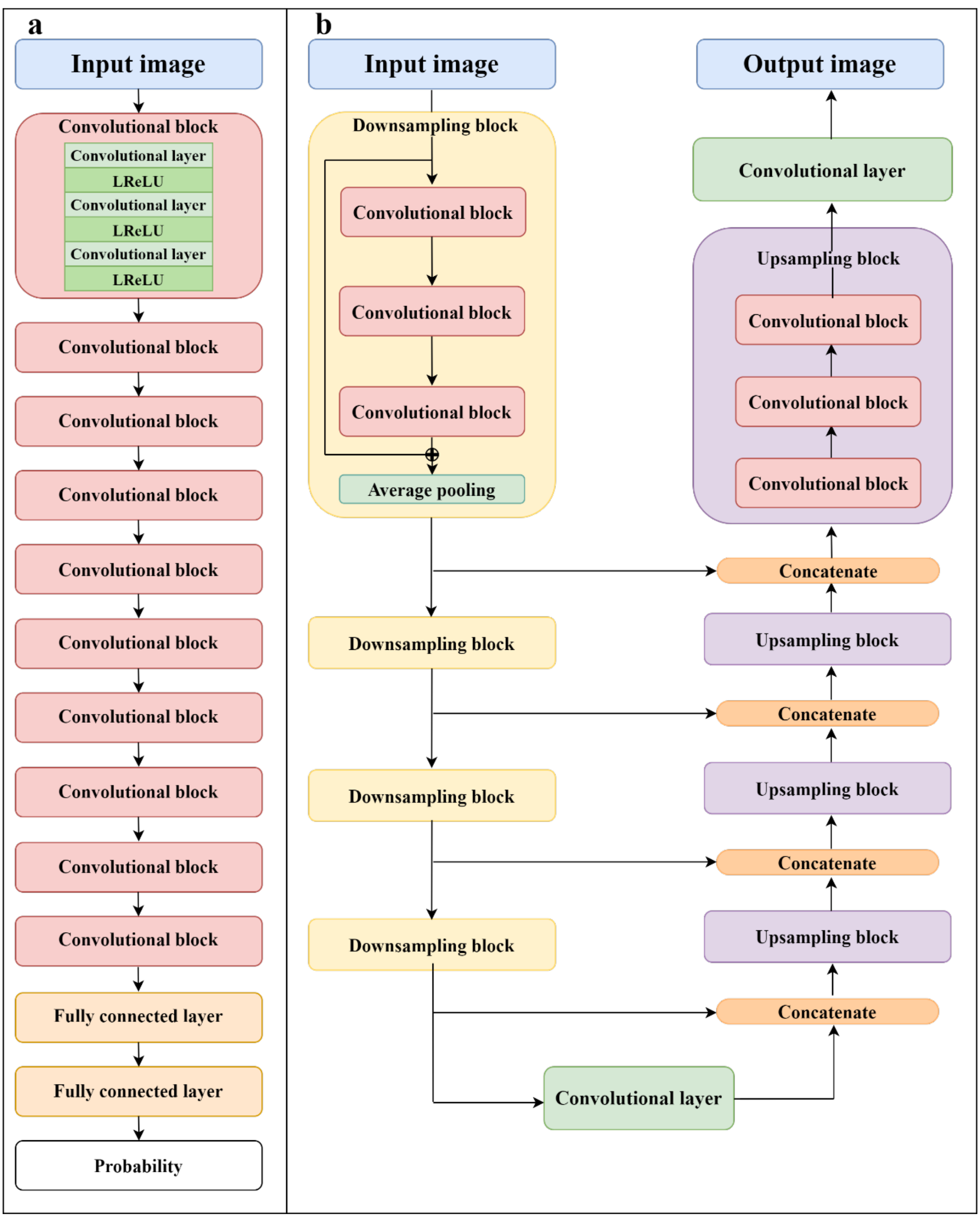

**Figure 1.** The models used in this research. (**a**) The discriminator model (input image can be an HQ ground truth image or an HQ generated image), (**b**) The generator model based on a U-NET architecture (input image is an LQ image, and the output is an HQ generated image).

Altogether, training was performed on a set of 4,800 image pairs, all with dimensions of 256 x 256 pixels. The rest of the data was left as a testing and validation set (120 images). Our GAN architecture is designed to start with random initializations and is optimized using an Adam optimizer, starting at a learning rate of $10^{-5}$ for both the discriminator and the generator. It is also worth mentioning that the GAN network continuously switches between training the generator given the discriminator and updating the discriminator by keeping the generator unchanged. In our case, the generator model was saved every 10,000 epochs. Furthermore, during each iteration, the model was validated on the validation image set, and SSIM and PSNR scores were stored in memory, allowing the best model to be saved. This architecture was implemented using Python version 3.8.10, tensorflow version 2.13.0, numpy version 1.23.5, and skimage version 0.19.3. The training was performed using Google CoLaboratory (a cloud-based platform providing access to the graphics processing units). Here, a T4 GPU was used, and the model was trained for approximately 500,000 iterations. The details of the architecture can be found on the GitHub page. After saving the best model, we validated our solution on unpaired images, allowing us to assess the model's ability to generalize the learned features from the HQ images.

## 3. Results

In the following sections, we present representative results from network training, obtained on the test subset of the data. The metrics we used for image comparison are mean squared error (MSE), normalized root mean squared error (NRMSE), structural similarity index (SSIM), and peak signal-to-noise ratio (PSNR). These metrics allowed us to assess the fidelity of the generated images compared to the ground-truth images.

To analyze the effectiveness of our network, we processed images of microtubule filaments in mouse embryonic fibroblasts (MEFs) using our GAN model as well as BatchDeconvolution [42] software (**Figure 2**). Deconvolution reduced noise in the LQ wide-field images to some extent (**Figure 2d, h, m**). Nevertheless, our architecture handles this problem in a much more refined way. The microtubular structure is better defined in comparison to the LQ or LQ deconvoluted images. Interestingly, in some cases, our network was able to recover fine details of the microtubular network by reducing excess noise, thereby showing a clearer distribution of microtubular fibers and, in some regions, giving the impression of higher image quality than the HQ ground-truth image (**Figure 2xi**).

We compared the MSE, NRMSE, SSIM, and PSNR metrics for representative images of MEF cells acquired with confocal microscopy and wide-field fluorescence microscopy, as well as images generated by the network (**Figures 3-5, Table 1**). In both **Figures 3** and **4,** the proposed model was able to restore very intricate details from the cell structures, including actin filaments, microtubules, and even the internal structure of chromatin inside the nuclei. The images in **Figure 3** prove the ability of the network to transfer the high-quality features from the HQ ground truth images to the LQ ones. It is clear in panels **b** and **c**, as well as **h** and **i**, that the network was able to recover subtle information about the microtubule structure, which is difficult to distinguish in LQ images. Furthermore, in some cases (as shown in Figure 3), GAN-generated images exhibit greater detail than the HQ ground-truth images. This is especially visible in panels **f** and **i** of the GAN results, where the microtubule structures are more prominent than in the confocal ones. Moreover, a clear reduction in noise is evident in panel **f** when comparing confocal and generated images. The metrics we used for comparison (**Table 1**) indicate that the network was able to learn the features of the HQ modality and generated HQ images with high fidelity. Nonetheless, the model sometimes struggled to reconstruct actin stress fibers (**Figure 3a-c**, indicated by the white arrow). However, its denoising ability was able to extract information about the microtubule structure from the LQ images (**Figure 3e** and **h**), making them clearly visible (**Figure 3d-i)**.

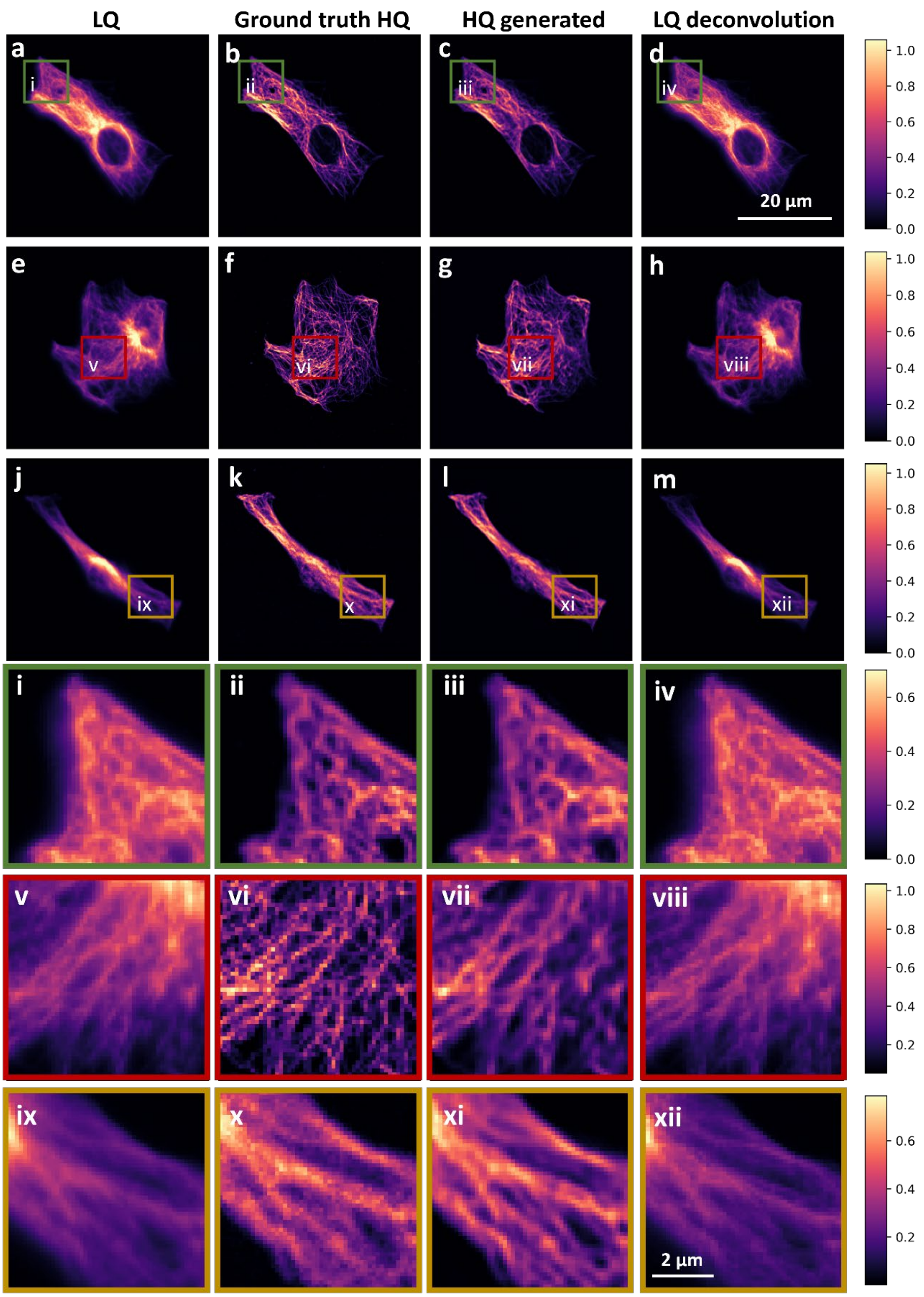

**Figure 2.** Comparison of LQ, ground-truth HQ, generated HQ, and deconvolved images of microtubule networks. The color bar on the right corresponds to the normalized fluorescence intensity. Panels (**a – m**) show the LQ input images, ground truth HQ (confocal) images, generated HQ, as well as LQ deconvoluted images. The insets (**i–xii**) correspond to the frames of the respective colors in the upper figures.

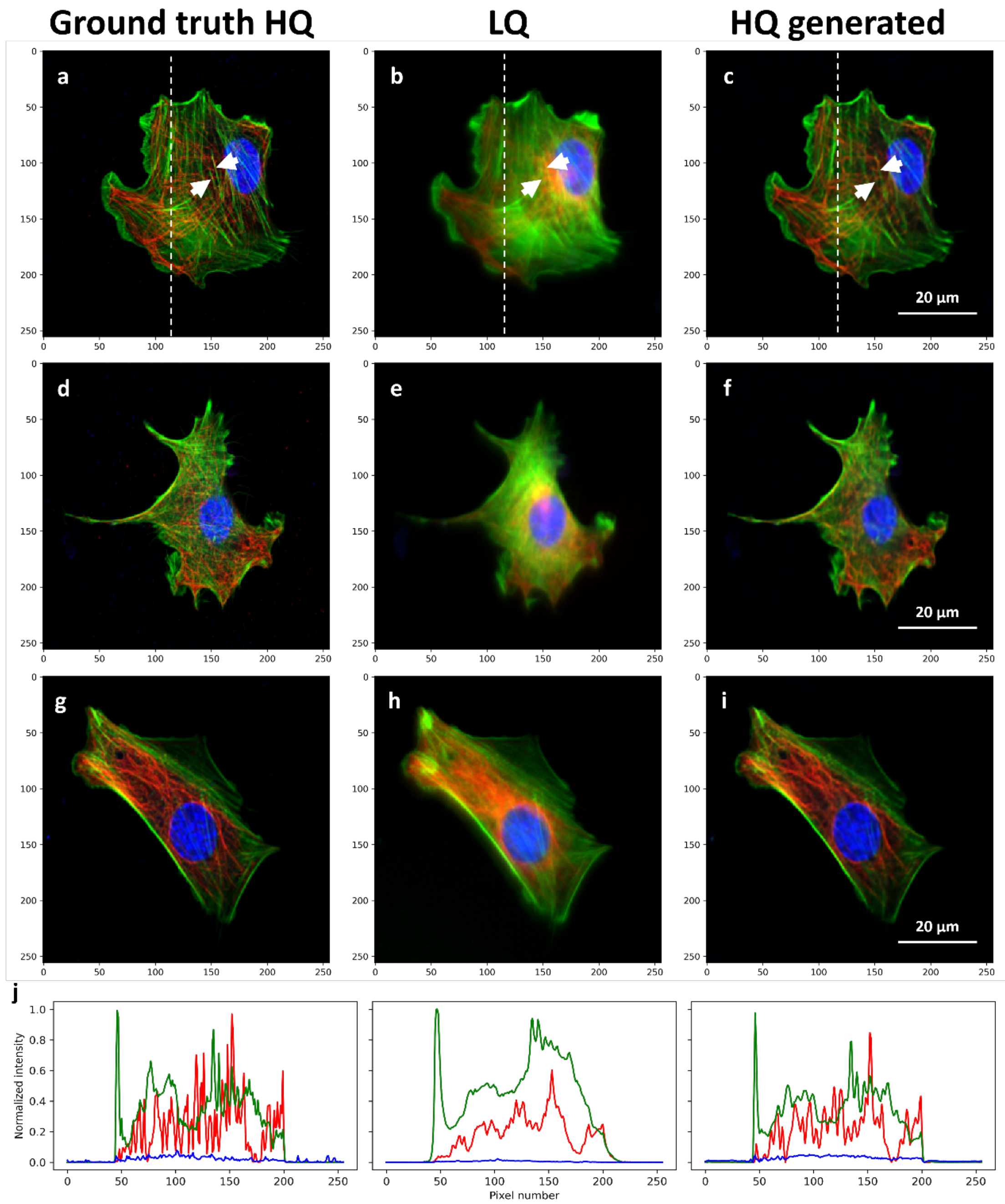

**Figure 3.** Exemplary images showing the network's ability to generate accurate HQ images from the LQ ones. **(a – i)** Comparisons of ground truth HQ (confocal), LQ (WWF), and HQ generated images. The white dashed lines in panels **a-c** correspond to the profiles shown in panel **j** (all the profiles are drawn over 128 pixels). Here, red represents microtubule fibers, green - actin filaments, and blue - nucleus. The white arrows point to the regions where the model struggled to reconstruct actin stress fibers. Numbers along the vertical and horizontal axes indicate the pixel number.

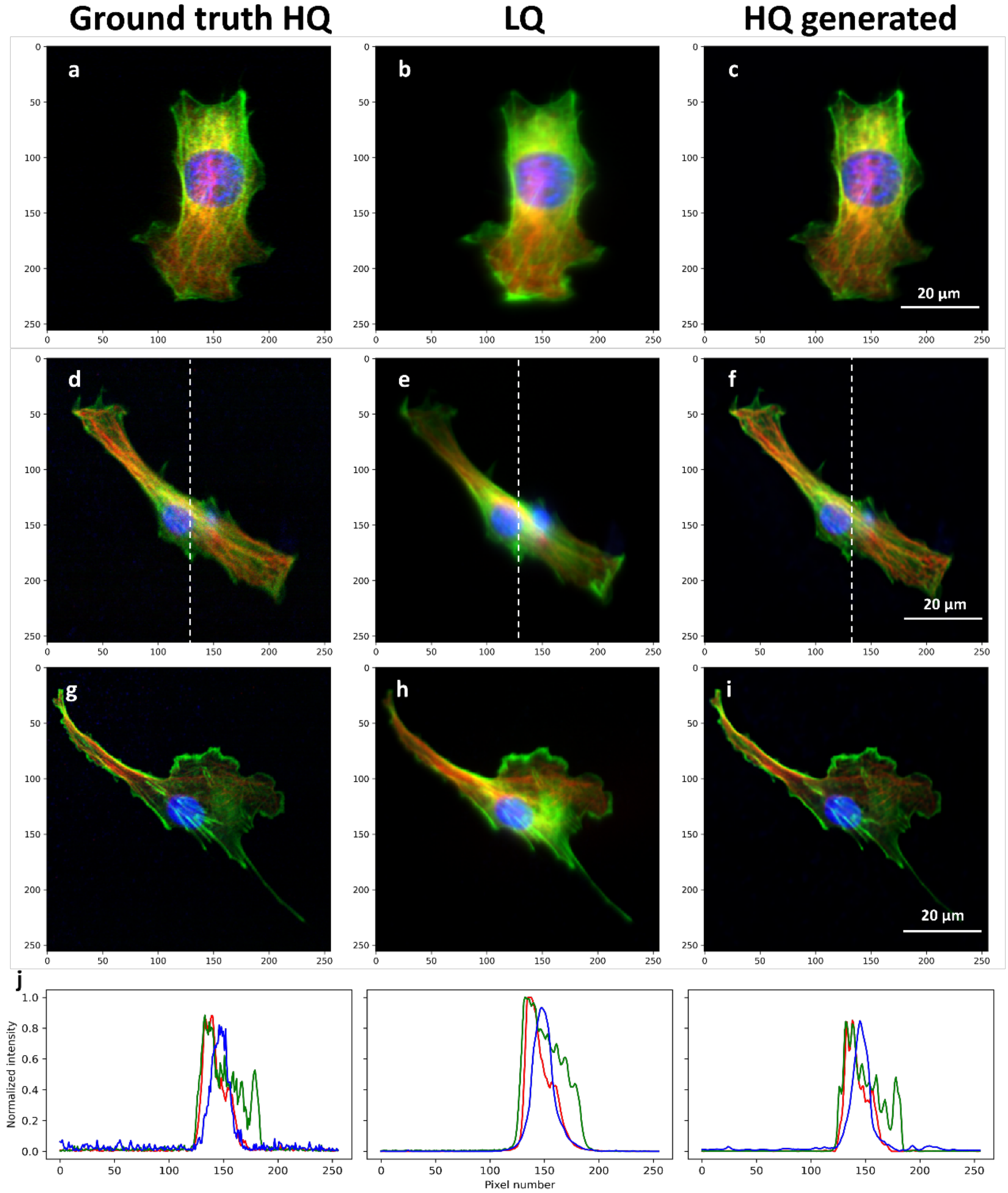

**Figure 4.** Exemplary images showing the network performance in generating HQ images from LQ Images. Panels **(a – i)** show comparisons of ground truth HQ (confocal), LQ (WWF), and HQ-generated images. Interestingly, the network was able to produce images with lower noise than the HQ ground-truth images. The white dashed lines in **d-f** correspond to the profiles shown in panel **j** (all the profiles drawn along 130 pixels). Here, red represents microtubule fibers, green - actin filaments, and blue - nucleus. Numbers along the vertical and horizontal axes indicate the pixel number.

**Table 1.** Metrics comparing images shown in Figures 2 and 3. Values of the comparisons between ground truth HQ and generated HQ images are bolded. MSE – mean squared error, NRMSE – normalized MSE, SSIM – structural similarity index, PSNR – peak signal-to-noise ratio.

| Comparison | | MSE | NRMSE | SSIM | PSNR |
|---|---|---|---|---|---|
| | a vs. b | 0.0106 | 0.7249 | 0.8142 | 19.73 |
| | a vs. c | 0.0011 | 0.2315 | 0.9283 | 29.65 |
| | b vs. c | 0.0092 | 0.5084 | 0.8292 | 20.35 |
| | d vs. e | 0.0077 | 0.6568 | 0.8320 | 21.15 |
| **Fig. 3** | d vs. f | 0.0010 | 0.2396 | 0.9271 | 29.91 |
| | e vs. f | 0.0070 | 0.5004 | 0.8512 | 21.55 |
| | g vs. h | 0.0066 | 0.6166 | 0.8464 | 21.77 |
| | g vs. i | 0.00051 | 0.1708 | 0.9452 | 32.92 |
| | h vs. i | 0.0064 | 0.4582 | 0.8375 | 21.91 |
| | a vs. b | 0.0037 | 0.3589 | 0.8517 | 24.34 |
| | a vs. c | 0.00067 | 0.1437 | 0.9230 | 31.71 |
| | b vs. c | 0.0033 | 0.2852 | 0.8638 | 24.86 |
| | d vs. e | 0.0044 | 0.5002 | 0.6723 | 23.54 |
| **Fig. 4** | d vs. f | 0.00060 | 0.1840 | 0.8410 | 32.23 |
| | e vs. f | 0.0040 | 0.4701 | 0.8281 | 23.97 |
| | g vs. h | 0.0071 | 0.8052 | 0.7570 | 21.46 |
| | g vs. i | 0.0005 | 0.2156 | 0.9080 | 32.90 |
| | h vs. i | 0.0068 | 0.5548 | 0.8581 | 21.65 |

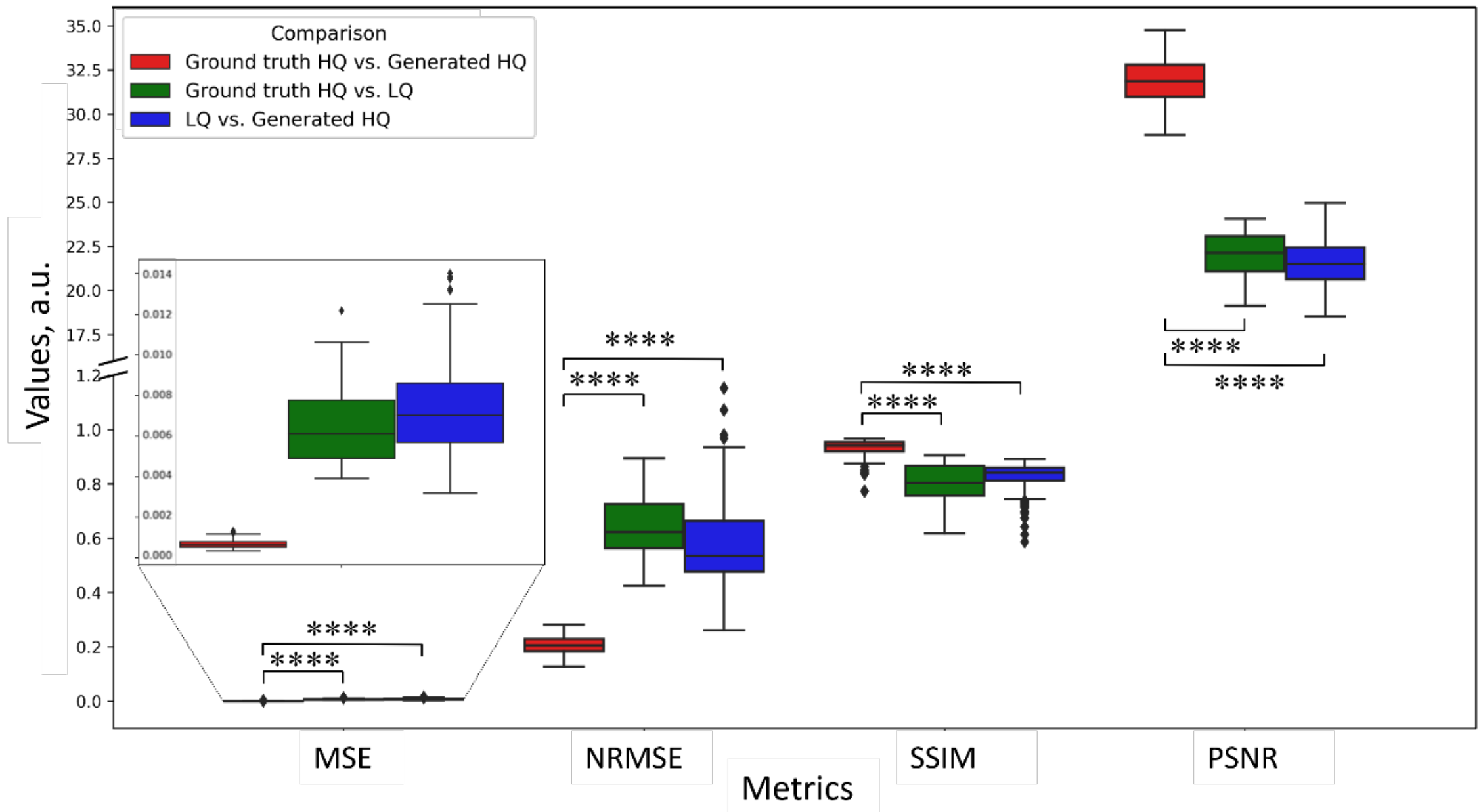

**Figure 5.** Box plots representing differences between the metrics calculated on the testing set. The line within a box represents the median value, the upper and lower box boundaries represent the first and the third quartile respectively, the whiskers represent a range of the data (1.5 times the interquartile range), and the points under or above the whiskers represent the outliers. **** $p < 0.0001$ (statistics calculated using non-parametric Kruskal-Willis ANOVA test, and *post hoc* Dunn's test).

Additionally, the final model was tested on images for which the ground-truth HQ images were unknown. The model was able to reconstruct the microtubular structure, which is not clearly visible in the LQ image (**Supplementary Figure 1**), demonstrating promising robustness to changes in microscopy systems over time, such as decreasing light source power or micro-damage to the microscope's optical elements. This procedure also allowed us to estimate the extent to which experimental conditions and/or microscope settings can influence the model's outcome.

## 4. Discussion

In this study, we demonstrated that a deep learning approach based on GAN architecture can be used for modality transfer between physically separate microscopy systems. Using paired wide-field fluorescence and confocal datasets acquired on independent instruments, our model generated images that more closely resembled the confocal reference data than the original wide-field inputs and outperformed conventional deconvolution (**Figure 2**). The restored images showed improved structural definition and reduced noise, with a visual appearance closer to that of the confocal reference images. These improvements were evident not only in representative examples of microtubule reconstruction (**Figure 2**), but also across other cellular features, including actin filaments and nuclear structure, where the model recovered details that were difficult to resolve in the original wide-field images (**Figures 3** and **4**). Quantitative comparisons further supported these observations, with generated images showing consistently better agreement with the confocal reference data than the low-quality inputs, as reflected by SSIM, PSNR, MSE, and NRMSE values (**Table 1**, **Figure 5**).

The applicability of this approach depends on the relationship between the training data and the conditions under which new data are acquired. Imaging settings, system state, staining quality, sample preparation, and biological sample type may all affect model performance. Accurate spatial alignment of the paired training images is also critical, because any mismatch between modalities can reduce the quality of the learned transformation and limit model performance [48]. Accordingly, careful validation is required when extending the model beyond the domain represented in the training set, even though the results obtained on previously unseen data acquired under slightly altered conditions indicate promising robustness (**Supplementary Figure 1**).

Importantly, the value of this work lies not only in image enhancement itself, but in demonstrating that image quality can be computationally transferred between independent microscopy systems. Here, we demonstrated this approach using wide-field fluorescence and confocal microscopy as a proof of principle. However, we believe that the same general strategy could be extended to other cross-modality setups, in which one system provides faster acquisition or greater accessibility, while the other provides higher image quality or more informative readouts. Possible examples include confocal – STED [22] or wide-field fluorescence – STORM [49,50] combinations, where the practical gains in acquisition time, instrument availability, and cost efficiency could be even greater than in the wide-field/confocal case.

This suggests a practical workflow in which large-scale image acquisition can be performed on fast and accessible platforms, while advanced, lower-throughput instruments are reserved for targeted validation and reference data generation. Such a workflow is particularly relevant in high-throughput and high-content imaging, where acquisition time on advanced systems often becomes a major limitation. In this way, once a paired database has been generated, the screening portion of the study could be performed on a fast system, while only the most relevant findings would be validated on a high-end instrument (**Figure 6**).

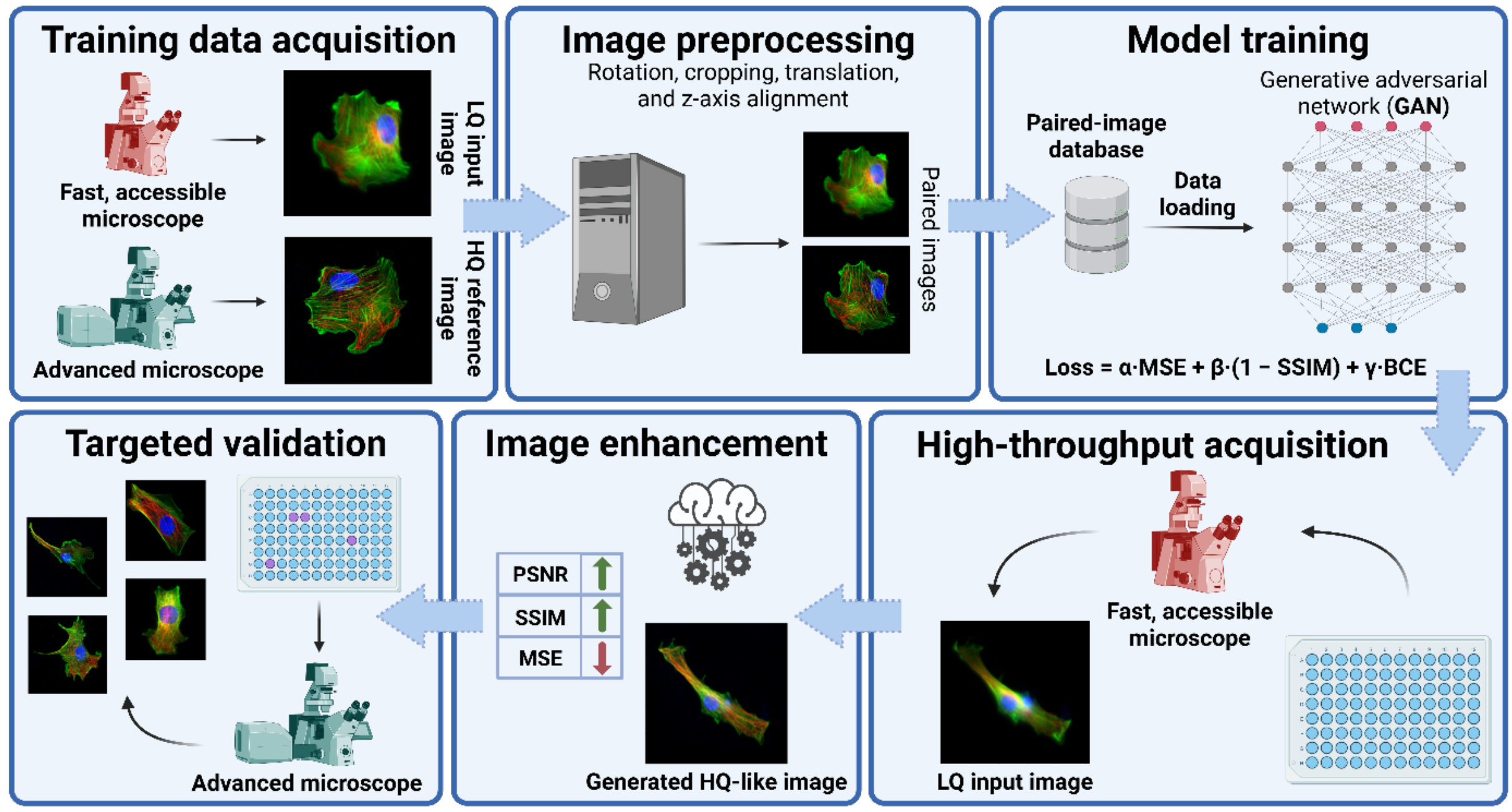

**Figure 6.** Proposed workflow for cross-instrument deep learning-enabled modality transfer. Paired low-quality (LQ) and high-quality (HQ) reference images are acquired using a fast, accessible microscope and an advanced microscope, respectively. The images are aligned and preprocessed before being used to train the generative adversarial network. After training, high-throughput acquisition can be performed on the fast microscope, and the resulting LQ images are transformed into HQ-like images. Selected results can then be further validated on the advanced microscope. This workflow reduces the time required on the advanced system while simultaneously allowing it to support larger high-throughput studies and more users.

This is particularly valuable from the perspective of imaging core facilities and shared infrastructure [51,52]. High-end microscopes are expensive to purchase, maintain, and operate, and their capacity is often limited. If part of the imaging workflow can be shifted to simpler or more available systems, while image quality is computationally enhanced using models trained against a higher-end reference platform, this could help reduce pressure on premium instruments, improve their availability, and lower the cost and time associated with large-scale imaging experiments. In such a workflow, advanced systems would still remain essential, but their use could be focused on tasks where they provide the greatest value.

A second possible use of this approach is to support groups with limited routine access to advanced imaging systems, including those in low- and middle-income countries. In such cases, the paired training data would still need to be generated using both the local system and a higher-end reference microscope. Yet, the reference microscope could be located at another institution or even in another country. Once the paired database has been established, the local system could be used for larger-scale acquisition within the validated range of the model. This could make lower-cost or older systems more useful and help expand access to high-quality imaging in laboratories with limited resources.

## 5. Conclusions

To our knowledge, this is the first work showing deep learning-based image enhancement between datasets acquired on physically separate imaging systems. Although demonstrated here using wide-field and confocal microscopy, the main significance of this work lies in establishing a broader proof of principle for computational image-quality transfer across independent systems. The same general strategy could be adapted to other modality pairs in which one system offers higher speed, lower cost, or greater accessibility, while the other provides higher image quality or more informative readouts. As such, this approach could support more scalable high-throughput and high-content imaging and enable more efficient use of advanced microscopy platforms within larger imaging workflows. Accordingly, this framework supports a model in which well-equipped and well-staffed shared imaging facilities can provide access to advanced equipment and expertise; simultaneously, it offers a way to reduce instrument time, even for high-throughput studies, thereby making these techniques accessible to more users.

**Acknowledgments**

This research was funded by the Priority Research Area DigiWorld under the Excellence Initiative – Research University program at the Jagiellonian University in Kraków (to ZR and ZB) and The Polish National Science Centre PRELUDIUM supported this work grant no. 2018/31/N/NZ3/02031 (to ZB), and supported by the Research support module as part of the Excellence Initiative – Research University program at the Jagiellonian University in Kraków no. RSM/60/PD (to DP). Figure 6 was created in BioRender.

**Data availability statement**

The data that support the findings of this study are available from the corresponding author upon reasonable request. The GAN software is available to download from the GitHub repository: github.com/panekdominik/Projects/tree/main/GAN_imge_superresolution.

**Authors contribution**

D.P. wrote the software. C.R., M.Sz., and J.S. prepared a paired image database and additional validation images. D.P. tested the algorithm and prepared figures. Z.B., Z.R., and D.P. designed the study. D.P. prepared the first draft of the manuscript. D.P. and Z.B. wrote the final version of the manuscript. Z.B., Z.R., and K.M. supervised the study. All authors contributed to the discussion and accepted the final version of the manuscript.

The authors declare no competing interests.

**Supplementary materials**

1. Supplementary Figure 1
2. Supplementary Note 1

**Declaration of generative AI and AI-assisted technologies in the manuscript preparation process**

During the preparation of this work the authors used ChatGPT (OpenAI) and Grammarly in order to improve language, readability, and phrasing of the manuscript text. After using this tool/service, the authors reviewed and edited the content as needed and take full responsibility for the content of the published article.

**Democratizing Advanced High-Throughput Imaging via Cross-Instrument Deep Learning-Enabled Modality Transfer**

Dominik Panek[a, b]*, Carina Rząca[a, b], Maksymilian Szczypior[b,c], Joanna Sorysz[d,#], Krzysztof Misztal[e], Zbigniew Baster[b, f]*, Zenon Rajfur[b]*

[a] Doctoral School of Exact and Natural Sciences, Jagiellonian University, ul. Łojasiewicza 11, 30-348 Kraków, Poland.
[b] Department of Molecular and Interfacial Biophysics, Faculty of Physics, Astronomy and Applied Computer Science, Jagiellonian University, ul. Łojasiewicza 11, 30-348 Kraków, Poland.
[c] Undergraduate program in Biophysics, Faculty of Physics, Astronomy and Applied Computer Science, Jagiellonian University, ul. Łojasiewicza 11, 30-348 Kraków, Poland.
[d] Undergraduate program in Biomedical Engineering, Faculty of Electrical Engineering, Automatics, Computer Science and Biomedical Engineering, AGH University of Science and Technology, Al. A. Mickiewicza 30, Building B-1, 30-059 Kraków, Poland.
[e] Division of Computational Mathematics, Faculty of Mathematics and Computer Science, Jagiellonian University, 6 Łojasiewicza St., Kraków, Poland.
[f] Laboratory for Cell and Tissue Engineering, Department of Biomedical Engineering, Eindhoven University of Technology, 5600 MB Eindhoven, The Netherlands.
[#] Current address: German Research Center for Artificial Intelligence (DFKI) and RPTU Kaiserslautern-Landau, Kaiserslautern, Germany.
*Author to whom correspondence should be addressed: dominik.panek@doctoral.uj.edu.pl, z.j.baster@tue.nl, zenon.rajfur@uj.edu.pl.

# Supplementary materials

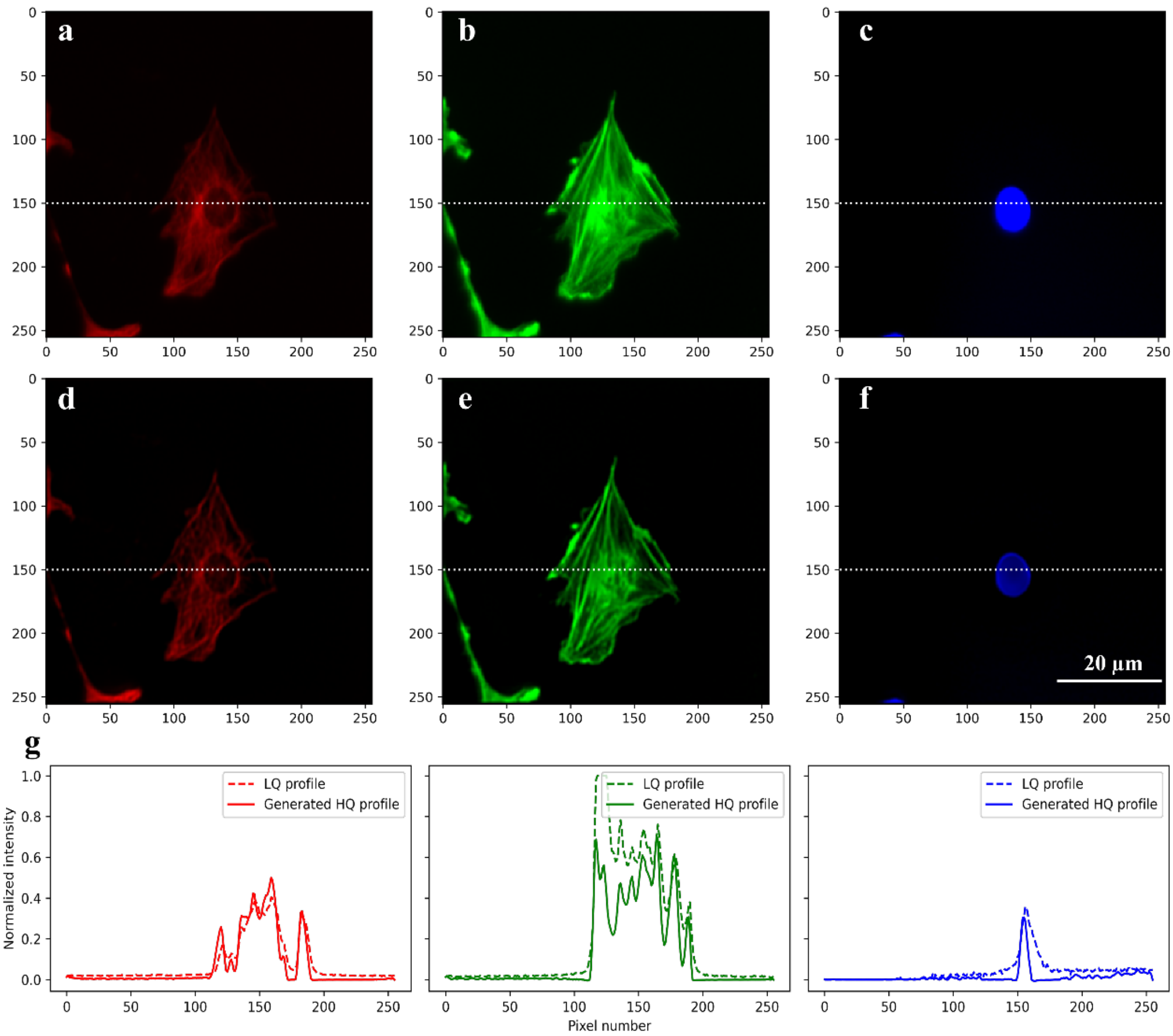

**Supplementary Figure 1.** Representative results of model performance on an unseen wide-field fluorescence microscopy image. The WFF image was taken under slightly different conditions than the training and testing data. **(a-c)** WFF images, **(d – f)** generated HQ image, **(d)** cross sections through the images (plotted according to the horizontal line in the **(f)** image). Dashed and solid lines represent cross-sections through the WWF and HQ generated images, respectively. Red represents microtubule fibers, green - actin filaments, and blue - nucleus. Numbers along the vertical and horizontal axes indicate the pixel number.

**Supplementary Note** 1 - **Ablation Studies**

To adjust the loss function parameters (**Equation 6**) the ablation studies were performed to compare different values of $\alpha$, $\beta$, and $\gamma$ parameters and how they can impact the final result of quality transfer.

We performed an ablation over the loss-term weights α, β, γ to determine a robust trade-off between the loss functions. **Supplementary Table N1** and **Supplementary Figure N1** summarize representative combinations. For each setting we report validation SSIM, PSNR and MSE and visually inspect the results. To further ensure that the contribution of each component was properly balanced, the MSE and SSIM weights were systematically examined in both ascending and descending order while keeping the adversarial (BCE) term constant. The symmetry of these trends confirmed the stability of the model's response to weight variation and helped us identify an optimal region where perceptual quality and structural integrity are jointly preserved.

Based on these results, we selected **α = 0.01, β = 0.1, γ = 1** as the final setting because it maximized SSIM and PSNR while avoiding visually apparent hallucinated structures.

**Supplementary Table N1.** Ablation study evaluating different weight combinations of the loss-function components **α** (MSE), β (SSIM), and γ (BCE). For each configuration, the relative contribution of each loss term and the resulting performance metrics (PSNR, SSIM, and MSE) are reported.

| α | β | γ | sum | share α [%] | share β [%] | share γ [%] | PSNR | SSIM | MSE |
|---|---|---|---|---|---|---|---|---|---|
| **0.01** | **0.1** | **1** | **1.11** | **1** | **9** | **90** | **31.86896** | **0.94131** | **0.0007** |
| 0.04 | 0.1 | 1 | 1.14 | 4 | 9 | 88 | 28.93127 | 0.91642 | 0.0013 |
| 0.1 | 0.1 | 1 | 1.2 | 8 | 8 | 83 | 27.63457 | 0.89668 | 0.0017 |
| 0.2 | 0.1 | 1 | 1.3 | 15 | 8 | 77 | 27.41727 | 0.90552 | 0.0018 |
| 0.5 | 0.1 | 1 | 1.6 | 31 | 6 | 63 | 27.05092 | 0.89989 | 0.002 |
| 0.7 | 0.1 | 1 | 1.8 | 39 | 6 | 56 | 25.10338 | 0.88216 | 0.0031 |
| 0.9 | 0.1 | 1 | 2.0 | 45 | 5 | 50 | 25.23878 | 0.88679 | 0.003 |
| 0.1 | 0.04 | 1 | 1.14 | 9 | 4 | 88 | 31.86899 | 0.88727 | 0.0007 |
| 0.01 | 0.2 | 1 | 1.21 | 1 | 17 | 83 | 27.30984 | 0.89111 | 0.0019 |
| 0.01 | 0.5 | 1 | 1.51 | 1 | 33 | 66 | 27.27799 | 0.89283 | 0.0019 |
| 0.01 | 0.7 | 1 | 1.71 | 1 | 41 | 58 | 26.59114 | 0.90108 | 0.0022 |
| 0.01 | 0.9 | 1 | 1.91 | 1 | 47 | 52 | 25.29531 | 0.93031 | 0.003 |
| 0.04 | 0.04 | 1 | 01.0 | 4 | 4 | 93 | 31.86898 | 0.93012 | 0.0007 |
| 0.1 | 0.1 | 1 | 1.2 | 8 | 8 | 83 | 27.65196 | 0.86543 | 0.0017 |
| 0.2 | 0.2 | 1 | 1.4 | 14 | 14 | 71 | 27.5907 | 0.86406 | 0.0017 |
| 0.5 | 0.5 | 1 | 2.0 | 25 | 25 | 50 | 25.07718 | 0.88216 | 0.0031 |
| 0.7 | 0.7 | 1 | 2.4 | 29 | 29 | 42 | 24.82647 | 0.91403 | 0.0033 |
| 0.9 | 0.9 | 1 | 2.8 | 32 | 32 | 36 | 24.69398 | 0.91159 | 0.0034 |

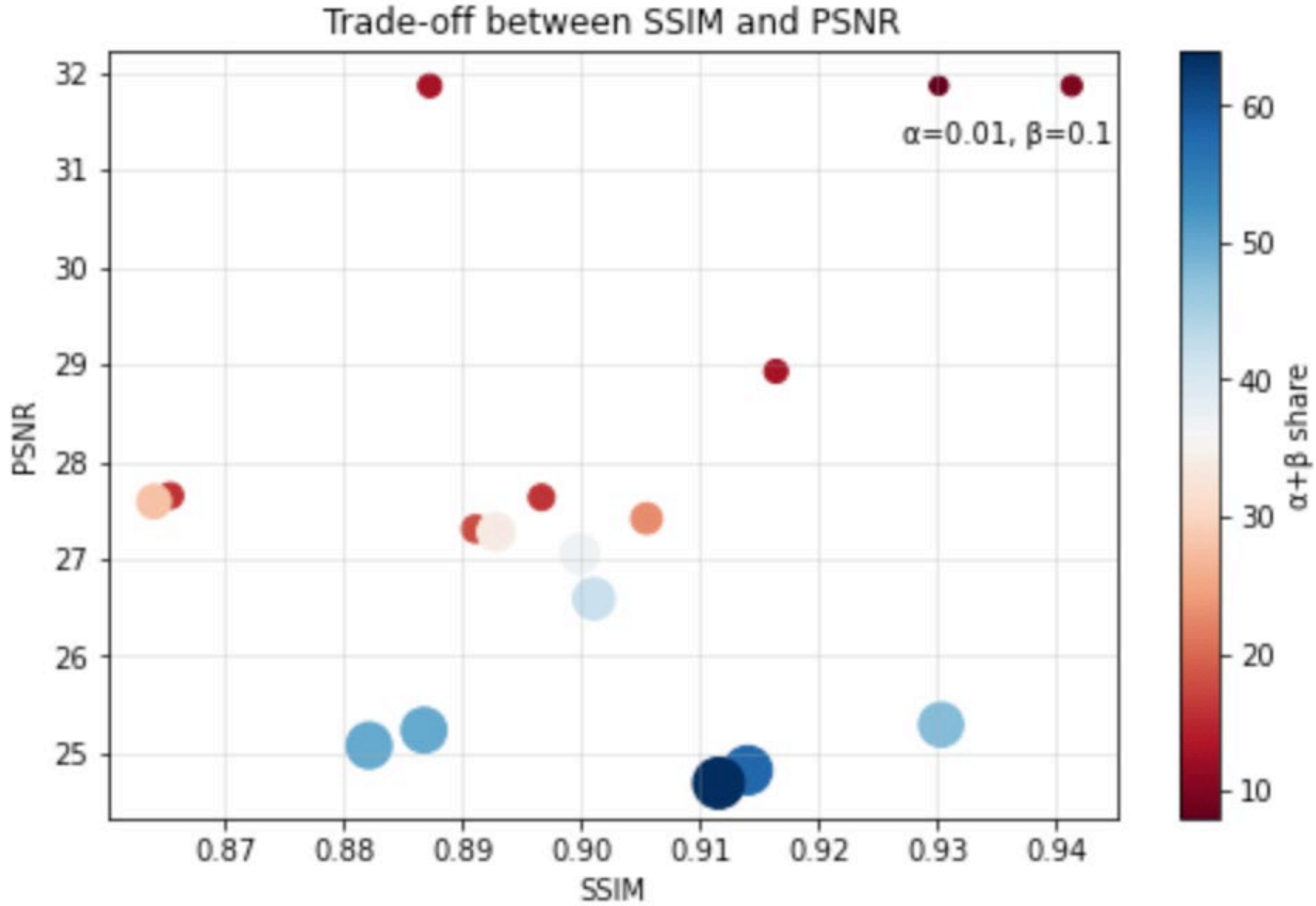

**Supplementary Figure N1.** Trade-off between SSIM and PSNR obtained for different combinations of loss-function weights. Marker color indicates the relative contribution of the reconstruction terms (**α+ β**). The selected configuration is highlighted.